\documentstyle{article}
\begin{document}
\begin{center}
{\bf{An Alternative Method for Solving a Certain Class of Fractional Kinetic Equations}}\\[0.5cm]

R.K. SAXENA\\
Department of Mathematics and Statistics, Jai Narain Vyas University\\
Jodhpur-342 004, India\\[0.3cm]

A.M. MATHAI\\
Department of Mathematics and Statistics, McGill University\\
Montreal, Canada H3A 2K6\\
and\\
Centre for Mathematical Sciences, Pala Campus, Pala-686 574, Kerala, India\\[0.3cm]

H.J. HAUBOLD\\
Office for Outer Space Affairs, United Nations\\
P.O.Box 500, A-1400 Vienna, Austria\\
and\\
Centre for Mathematical Sciences, Pala Campus, Pala-686 574, Kerala, India\\[0.3cm]

\end{center}
\noindent
{\bf Abstract.} An alternative  method for solving the fractional kinetic equations solved earlier by Haubold and Mathai (2000) and Saxena et al. (2002, 2004a, 2004b) is recently  given by Saxena and Kalla (2007). This method can also be applied in solving more general fractional kinetic equations than the ones solved by the aforesaid authors. In view of the usefulness and importance of the kinetic equation in certain physical problems governing reaction-diffusion in complex systems and anomalous diffusion, the authors present an alternative simple method for deriving the solution of the generalized forms of the fractional kinetic equations solved by the aforesaid authors and Nonnenmacher and Metzler (1995). The method depends on the use of the Riemann-Liouville fractional calculus operators. It has been shown by the application of  Riemann-Liouville  fractional integral operator and  its interesting properties, that the solution of the given fractional kinetic equation can be obtained in a straight-forward manner. This method does not make use of the Laplace transform.
\section{Introduction}
The paper deals with the essential problem related to applications of  Mittag-Leffler  function and Riemann-Liouville fractional calculus operators to fractional order kinetic equations arising in modeling physical phenomena, governing diffusion in porous media and relaxation processes. As such it reveals the important role of these tools in applications of fractional calculus. The results are interesting and useful  for wide range of applied scientists dealing with fractional order differential  and fractional order integral equations. In a series of papers the authors have demonstrated the use of integral transforms in the solution of certain fractional kinetic equations (2002, 2004a 2004b), reaction-diffusion equations  (2006a, 2006b), and fractional differential equations governing nonlinear waves (2006c, 2006d ). In the present paper it is shown by the application of  Riemann-Liouville fractional calculus operators and its interesting properties that the given fractional kinetic equations can be easily solved. Fractional kinetic equations are studied by Zaslavsky (1994), Saichev and Zaslavsky (1997), Gloeckle and Nonnenmacher (1991), and Saxena, Mathai and Haubold (2002, 2004a, 2004b) due to their importance in the solution of certain applied problems governing reaction and relaxation in complex systems and anomalous diffusion. The use of fractional kinetic equations  in many  problems arising in science and engineering  can  be found in  the monographs by Podlubny (1999), Hilfer (2000), and Kilbas, Srivastava and Trujillo  (2006) and the various papers given  therein. 
      The Mittag-Leffler functions naturally occur as a solution of fractional order differential equation or a fractional order integral equation. Mittag-Leffler (1903) defined this function, known as Mittag-Leffler function in the literature, in terms of the power series 
\begin{equation}
E_\alpha(z):=\sum^\infty_{k=0}\frac{z^k}{\Gamma(\alpha k+1)};\;\;(\alpha \in C, Re(\alpha)>0).
\end{equation}                      		
This function is generalized by Wiman (1905) in the form
\begin{equation}
E_{\alpha, \beta}(z):=\sum^\infty_{k=0}\frac{z^k}{\Gamma(\alpha k+\beta)}, (\alpha, \beta \in C, Re(\alpha)>0, Re(\beta) >0).
\end{equation}
According to Dzherbashyan (1966, p.118), both the functions defined by the equations (1) and (2) are entire functions of order $\rho=1/\alpha$ and type $\sigma =1$.  A comprehensive detailed account of these functions is available from the monographs of Erd\'{e}lyi, Magnus, Oberhettinger and Tricomi (1995, Chapter 18) and Dzherbashyan (1966, Chapter 2). 
       The Riemann-Liouville operators of fractional calculus are defined in the books by Miller and Ross (1993), Oldham and Spanier (1974), Podlubny (1999) and Kilbas, Srivastava and Trujillo (2006) as
\begin{equation}
_\alpha D_t^{-\nu} N(t):= \frac{1}{\Gamma(\nu)}\int^t_\alpha (t-u)^{\nu-1}\;N(u)du, Re(\nu)>0, t>a
\end{equation}        
with $_aD_t^0 N(t) = N(t),$  and
\begin{equation}
_aD_t^\mu N(t):=\frac{d^n}{dt^n}(_aD_t^{\mu-n} N(t)), Re(\mu)>0, n-\mu>0.
\end{equation}
By virtue of the definitions (3), it is not difficult to show that 
\begin{equation}
_aD_t^{-\nu}(t-a)^{\rho-1}=\frac{\gamma(\rho)}{\Gamma(\rho+\nu)}(t-a)^{\rho+\nu-1},
\end{equation}
where $Re(\nu)>0, Re(\rho)>0; t>a.$
Also from  (Podlubny, 1999, p.72, eq.(2.117)), we have
\begin{equation}
_aD_t^\nu(t-a)^{\rho-1}=\frac{\gamma(\rho)}{\Gamma(\rho-\nu)}(t-a)^{\rho-\nu-1},
\end{equation}
where $Re(\nu)>0, Re(\rho)>0, t>a.$ When $\rho=1$ (6) reduces to an interesting formula
\begin{equation}
_aD_t^{-\nu} 1=\frac{1}{\Gamma(1-\nu)}(t-a)^{-\nu}, t>a; \nu \neq 1,2\ldots
\end{equation}						
which is a remarkable result in the theory of fractional calculus and indicates that the fractional derivative of a constant is not zero.
	
       We now proceed to derive and solve the fractional kinetic equations in the next section.
       
\section{Derivation of the fractional kinetic equation and its solution} 

       If we integrate the standard kinetic equation
\begin{equation}
\frac{d}{dt}N_i(t)=-c_iN_i(t), (c_i>0)
\end{equation}        
we obtain (Haubold and Mathai, 2000, p.58)
\begin{equation}
N(t)-N(a)=-c_i\;\;_aD_t^{-1}N_i(t),
\end{equation}	                                                                              
where $_aD_t^{-1}$ is the standard Riemann integral operator. Here we recall that in the original paper of Haubold and Mathai (2000), the number density of species, $N_i=N_i(t)$  is a function of time. Further we assume that $N_i(t=a)=N_a$   is the number density of species $i$ at time  $t=a$. If we drop the index $i$ in (9) and generalize it, we arrive at the fractional kinetic equation 
\begin{equation}
N(t)-N_a= -c^\nu\;_aD_t^{-\nu}N(t)
\end{equation}
Solution of (10). If we multiply both sides of (10) by $(-c^\nu)^m\;_aD_t^{-m\nu)}$, we obtain 
\begin{equation}
(-c^\nu)^m\;_aD_t^{-m\nu} N(t)-(-c^\nu)(-c^\nu)^m\;_aD_t^{-m\nu-\nu}\;N(t)=(-c^\nu)^m\;_aD_t^{-m\nu}N_a.
\end{equation}          
Now summing  up both sides of (2.4) for $m$  from $0$ to $\infty$, it yields 
\begin{eqnarray}
&&\sum^\infty_{m=0}(-c^\nu)^m\;_aD_t^{-m\nu}N(t)-\sum^\infty_{m=0}(-c^\nu)^{m+1}\;_aD_t^{-m\nu-\nu}N(t)\nonumber\\
&=& N_a\sum^\infty_{m=0}(-c^\nu)^m\;_aD_t^{-m\nu}1,
\end{eqnarray}
which on using the formula (5) yields 
\begin{eqnarray}
N(t)& = & N_a\sum^\infty_{m=0}(-c^\nu)^m[(t-a)^{m\nu}/\Gamma(m\nu+1)]\\                                                 
&=& N_aE_\nu[-c^\nu(t-a)^\nu], t>a  
\end{eqnarray}
       Thus we arrive at the following theorem:\par
\medskip
\noindent
{\bf Theorem 1.} If $Re(\nu)>0, Re(c)>0$ then there exists the  unique solution of the integral equation 
\begin{equation}
N(t)-N_a=-c^\nu\;_aD_t^{-\nu}(t),
\end{equation}
given by
\begin{equation} 
N(t)=N_aE_\nu(-c^\nu(t-a)^\nu), \;t>a
\end{equation}					
with the Mittag-Leffler function defined by (1).

When $a\rightarrow 0$, (16) reduces to the following  result given by Haubold and Mathai (2000, p.63):\par
\medskip
\noindent
{\bf Corollary 1.1.} If, $Re(c)>0$ then the unique  solution of the integral equation 
\begin{equation}
N(t) - N_0=-c^\nu\;_0D_t^{-\nu}(t),
\end{equation}	 			
is given by  
\begin{equation}
N(t)=N_0E_\nu(-c^\nu t^\nu).
\end{equation}
\medskip
\noindent
{\bf Remark 1.}	If we apply the operator $_aD_t^\nu$ from the left to (10) and make use of (7), we obtain the fractional differential equation 
\begin{equation}
_aD_t^\nu N(t)-N_a\frac{(t-a)^{-\nu}}{\Gamma(1-\nu)}=-c^\nu N(t),\;\;t>a
\end{equation}
whose solution is also given by (16).  When a tends to zero in (16), it reduces to  one obtained by Nonnenmacher and Metzler (1995, p.156) for the fractional relaxation equation with $c$ replaced by $1/c$.\par
\medskip
\noindent
{\bf Remark 2.}  The method adopted in deriving the solution of fractional kinetic equation (8) is similar to that used by Al-Saqabi and Tuan (1996) for solving differ integral equations.
\section {Theorem 2.} 
If $min\left\{Re(\nu),Re(\mu)\right\}>0, Re(c)>0,$ then there exists the unique solution of the integral equation 
\begin{equation}	 				`			
N(t)-N_at^{\mu-1}=-c^\nu\;_aD_t^{-\nu}(t),
\end{equation}
given by  
\begin{equation}
N(t)=N_a\Gamma(\mu)(t-a)^{\mu-1}\;E_{\nu,\mu}(-c^\nu(t-a)^\nu), t>a
\end{equation}														
where  $E_{\nu,\mu}(t)$is the  generalized Mittag-Leffler function defined by (2).\par
\medskip
\noindent
{\bf Solution of (20).} If  we multiply both sides of (20) by $(-c^\nu)^m\;\;_aD_t^{-m\nu}$, we obtain 
\begin{equation}
(-c^\nu)^m\;_aD_t^{-m\nu}N(t)-(-c^\nu)(-c^\nu)^m\;_aD_t^{-m\nu-\nu}N(t)= N_a(-c^\nu)^m\;_aD_t^{-m\nu}t^{\mu-1}.
\end{equation}
        
Now summing  up both sides of (22) for m from 0 to $\infty$, it yields 
\begin{equation}
\sum^\infty_{m=0}(-c^\nu)^m\;_aD_t^{-m\nu}N(t)-\sum^\infty_{m=0}(-c^\nu)^{m+1}\;_aD_t^{-m\nu-\nu}N(t)=N_a\sum^\infty_{m=0}(-c^\nu)^m\;_aD_t^{-m\nu}t^{\mu-1},
\end{equation}   
which on using the formula (5) gives 
\begin{eqnarray}
N(t)&=&N_a\Gamma(\mu)\sum^\infty_{m=0}(-c^\nu)^m[(t-a)^{m\nu}/\gamma(m\nu+\mu)]\\
&=& N_aE_{\nu,\mu}[-c^\nu(t-a)^\nu],\;t>a.
\end{eqnarray}
	
This completes the proof of  Theorem  2.\par
\medskip
\noindent
For $a=0$, (25) reduces to the following result given by  Saxena, Mathai and Haubold (2002, p.283).\par
\medskip
\noindent
{\bf Corollary 2.1.} If  min $\left\{Re(\nu), Re(\mu)\right\} >0, R(c)>0$ then the solution of the integral equation 
\begin{equation}
N(t)-N_0t^{\mu-1}=-c^\nu\;_0D_t^{-\nu}(t)
\end{equation}	 				`		
is given by  
\begin{equation}
N(t)=N_0\Gamma(\mu)t^{\mu-1}\;E_{\nu,\mu}(-c^\nu t^\nu),
\end{equation}													
where  $E_{\nu,\mu} (t)$is the  generalized Mittag-Leffler function defined by (2).\par
\bigskip
\noindent
{\bf References}\par
\smallskip
\noindent
Al-Saqabi, B.N., Tuan, V.K.: Solution of  a fractional differ integral equation. Integral Transforms and Special Functions {\bf 4}, 321-326 (1996)\par
\smallskip
\noindent
Dzherbashyan, M.M.: Integral Transforms and Representation of Functions in Complex Domain (in Russian). Nauka, Moscow (1966)\par
\smallskip
\noindent
Dzherbashyan, M.M.: Harmonic Analysis and Boundary Value Problems in the Complex Domain. Birkhaeuser-Verlag, Basel and London (1993)\par
\smallskip
\noindent
Erd\'{e}'lyi, A., Magnus, W., Oberhettinger, F., Tricomi, F.G.: Tables of Integral Transforms. Vol.{\bf 2}, McGraw-Hill, New York, Toronto, and London (1954)\par
\smallskip
\noindent
Erd\'{e}'lyi, A., Magnus, W., Oberhettinger, F., Tricomi, F.G.: Higher Transcendental Functions. Vol. {\bf 3}, McGraw-Hill, New York, Toronto, and London (1955)\par
\smallskip
\noindent
Gloeckle, W.G., Nonnenmacher, T.F.: Fractional integral operators and Fox function in the theory of viscoelasticity. Macromolecules {\bf 24} 6426-6434 (1991)\par
\smallskip
\noindent
Haubold, H.J., Mathai, A.M.: The fractional kinetic equation and thermonuclear functions. Astrophysics and Space Science {\bf 273} 53-63 (2000)\par
\smallskip
\noindent
Hilfer, R.: Applications of Fractional Calculus in Physics. World Scientific Publishing Co., New York (2000)\par
\smallskip
\noindent
Kilbas, A.A., Srivastava, H.M., Trujillo, J.J.: Theory and Applications of Fractional Differential Equations. North-Holland Mathematics Studies {\bf 204}, Elsevier, Amsterdam (2006)\par
\smallskip
\noindent
Mathai, A.M., Saxena, R.K.: The H-function with Applications in  Statistics and Other Disciplines. John Wiley and Sons Inc., New York, London and Sydney (1978)\par
\smallskip
\noindent
Miller, K.S., Ross, B.: An Introduction to Fractional Calculus and Fractional Differential Equations. Wiley and Sons, New York (1993)\par
\smallskip
\noindent
Mittag-Leffler, G.M.: Sur la nouvelle function. C.R. Acad. Sci., Paris, {\bf 137} 554-558 (1903)\par
\smallskip
\noindent
Nonnenmacher, T.F., Metzler, R.: On the Riemann-Liouville fractional calculus and some recent applications. Fractals {\bf 3} 557-566 (1995)\par
\smallskip
\noindent
Oldham, K.B., Spanier, J.: The Fractional Calculus: Theory and Applications of Differentiation and Integration of Arbitrary Order. Academic Press, New York (1974)\par
\smallskip
\noindent
Podlubny, I.: Fractional Differential Equations. Academic Press, San Diego (1999)\par
\smallskip
\noindent
Saichev, A., Zaslavsky, M.: Fractional kinetic equations: solutions and applications, Chaos {\bf 7} 753-764 (1997)\par
\smallskip
\noindent
Saxena, R.K., Kalla, S.L.: 2007, On the solution of certain fractional kinetic equations. Accepted for publication in Applied Mathematics and Computation (2007)\par
\smallskip
\noindent
Saxena, R.K., Mathai, A.M., Haubold, H.J.: On fractional kinetic equations. Astrophysics and Space Science {\bf 282} 281-287 (2002)\par
\smallskip
\noindent
Saxena, R.K., Mathai, A.M., Haubold, H.J.: On generalized fractional kinetic equations. Physica A {\bf 344} 653-664 (2004a)\par
\smallskip
\noindent
Saxena, R.K., Mathai, A.M., Haubold, H.J.: Unified fractional kinetic equation and a fractional diffusion equation. Astrophysics and Space Science {\bf 290} 299-310 (2004b)\par
\smallskip
\noindent
Saxena, R.K., Mathai, A.M., Haubold, H.J.: Fractional reaction-diffusion equations. Astrophysics and Space Science {\bf 305} 289-296 (2006a)\par
\smallskip
\noindent
Saxena, R.K., Mathai, A.M., Haubold, H.J.: Solution of generalized fractional reaction-diffusion equations. Astrophysics and Space Science {\bf 305} 305-313 (2006b)\par
\smallskip
\noindent
Saxena, R.K., Mathai, A.M., Haubold, H.J.: Reaction-diffusion systems and nonlinear waves. Astrophysics and Space Science {\bf 305} 297-303 (2006c)\par
\smallskip
\noindent
Saxena, R.K., Mathai, A.M., Haubold, H.J.: Solution of fractional reaction-diffusion equations in terms of Mittag-Leffler functions. International Journal of Scietific Research {\bf 15} 1-17 (2006d)\par
\smallskip
\noindent
Wiman, A.: Ueber den Fundamentalsatz in der Theorie der Funktionen. Acta Mathematica {\bf 29} 191-201 (1905)\par
\smallskip
\noindent
Zaslavsky, G.M.: Fractional kinetic equation for Hamiltonian chaos. Physica D {\bf 78} 110-122 (1994)
\end{document}